
\input harvmac
\input tables

\noblackbox

\Title{\vbox{
\hbox{hep-ph/9402348, WIS-94/9/Feb-PH}}}
{\vbox{
\centerline{Naturally Small $x_s$?}}}
\bigskip
\centerline{Yosef Nir}
\bigskip
\centerline{\it Department of Particle Physics}
\centerline{\it Weizmann Institute of Science, Rehovot 76100, Israel}

\bigskip
\baselineskip 18pt

\centerline{\bf Abstract:}
\medskip
\noindent
Within the Standard Model, $x_s$ (the mixing parameter in the
$B_s-\bar B_s$ system) is constrained to the range
$7\leq x_s\leq 40$. We point out that if New Physics
contributes significantly to $x_d$ (the mixing parameter in the
$B_d-\bar B_d$ system), then $2\leq x_s\leq7$ is possible
without any fine-tuned cancellations between the Standard Model
and the New Physics contributions.

\Date{February 1994}

\newsec{$x_s$ in the Standard Model}
A measurement of $x_s$, the mixing parameter in the
$B_s-\bar B_s$ system, would be of much interest
\ref\AlLo{For recent reviews, see A. Ali and D. London,
 J. Phys. G19 (1993) 1069; DESY-93-022 (1993).}.
Within the Standard Model it will determine
the ratio $|V_{td}/V_{ts}|$ with relatively small
hadronic uncertainties. Furthermore, it will constrain
or may even discover New Physics.

Within the Standard Model, mixing in the $B_s-\bar B_s$
system is dominated by box diagrams with intermediate
top quarks. This gives
\eqn\xsSM{
x_s^{\rm SM}={G_F^2 m_W^2\over6\pi^2}\eta_{QCD}(y_tf_2(y_t))
(\tau_{B_s}m_{B_s})(B_{B_s}f_{B_s}^2)|V_{ts}V_{tb}|^2}
where $y_t=m_t^2/m_W^2$ and
\eqn\ftwo{
f_2(y)=1-{3\over4}{y(1+y)\over(1-y)^2}\left[1+{2y\over1-y^2}
\ln(y)\right].}
One way to calculate the Standard Model constraints on $x_s$
is to directly use \xsSM. The significant sources of
uncertainty are $m_t$
\ref\mtDzero{S. Abachi et al., the D0 collaboration,
 Fermilab Pub-94/004-E (1994).}
\ref\PDG{P. Langacker, in Review of Particle Properties, Phys. Rev
 D45 (1992) III.59.},
$f_{B_s}$
\ref\sommers{For a review of recent lattice results, see
R. Sommers, DESY 94-011 (1993).}
\ref\neubert{For a review of recent QCD sum-rule results, see
M. Neubert, SLAC-PUB-6263 (1993).}
and $\tau_{B_s}|V_{ts}|^2\approx\tau_b|V_{cb}|^2$
\ref\CLEOVcb{G. Crawford {\it et al.}, CLEO collaboration,
 CLEO-CONF-93-30 (1993).}:
\eqn\uncertaina{\eqalign{
m_t=&\ 165\pm35\ GeV,\cr
\sqrt{B_{B_s}}f_{B_s}=&\ 0.22\pm0.06\ GeV,\cr
\sqrt{\tau_b\over1.49\ ps}|V_{cb}|=&\ 0.037\pm0.007.\cr}}
(We use
\ref\Abada{A. Abada {\it et al.}, Nucl. Phys. B376 (1992) 172.}\
$B_B=1.16\pm0.07$. Note that this corresponds to the renormalization
group invariant definition of $B_B$. Accordingly, we use for
$\eta_{QCD}$ the value of $\eta_{2B}(m_t=150\ GeV)=0.5$
\ref\BJW{A. Buras, M. Jamin and P.H. Weisz, Nucl. Phys. B347 (1990) 491.}.)
Allowing these parameters to vary independently within
their $1\sigma$ ranges, we get
\eqn\SMdirect{
3\leq x_s\leq40.}
Another option is to use the theoretical
expression for the {\it ratio} $R\equiv x_s/x_d$,
\eqn\RSM{
R^{\rm SM}=\left({m_{B_s}\over m_{B_d}}{\tau_{B_s}\over\tau_{B_d}}
\right)\left({B_{B_s}f_{B_s}^2\over B_{B_d}f_{B_d}^2}\right)
\left|{V_{ts}\over V_{td}}\right|^2,}
together with the experimental value of $x_d$
to find $x_s$. The significant sources of uncertainty
here are $x_d$
\ref\moser{For a review of recent CLEO, ARGUS and LEP results,
 see H.-G. Moser, CERN-PPE/93-164 (1993).},
$f_{B_s}/f_{B_d}$
\ref\gjmsw{B. Grinstein {\it et al.}, Nucl. Phys. B380 (1992) 369;
A.F. Falk, Phys. Lett. B305 (1993) 268;
B. Grinstein, SSCL-Preprint-492 (1993).}\sommers\
and $|V_{ts}/V_{td}|$:
\eqn\uncertainb{\eqalign{
x_d=&\ 0.69\pm0.07,\cr
{B_{B_s}f_{B_s}^2\over B_{B_d}f_{B_d}^2}=&\ 1.35\pm0.15,\cr
|V_{ts}/V_{td}|=&\ 5\pm2.\cr}}
The upper bound on $|V_{td}|$ arises from CKM unitarity
(we used the recent CLEO range
\ref\CLEOVub{J. Bartelt {\it et al.}, Phys. Rev. Lett. 71 (1993) 4111.}\
$|V_{ub}/V_{cb}|=0.08\pm0.03$), and the lower bound from the
$x_d$ constraint.
This leads to $11\leq R^{\rm SM}\leq75$ and consequently
\eqn\SMRatio{
7\leq x_s\leq60.}
Combining the two methods \SMdirect\ and \SMRatio, we finally
get the Standard Model prediction
\eqn\SMfinal{
7\leq x_s\leq40.}

Experiments will soon be able to explore the region near the
lower bound in eq. \SMfinal. The question addressed in
this work is whether a violation of this bound is
likely in the presence of New Physics.

\newsec{$x_s$ beyond the Standard Model}
There are several possible ways in which New Physics
could lead to violation of the bounds in \SMfinal:
\item{(a)} The ratio $|V_{ts}/V_{td}|$ is outside
the bounds \uncertainb.
\item{(b)} There are significant new contributions to $x_s$.
\item{(c)} There are significant new contributions to $x_d$.

1. We would first like to argue that the first effect (a) is
not really of much significance. The lower bound in \SMfinal\
corresponds to the upper bound on $|V_{td}|$. This, as
mentioned above, is a result of CKM unitarity; therefore
it can only be violated in models where the quark sector
is extended beyond the three sequential generations of
the Standard Model. It was shown, however, in ref.
\ref\nisi{Y. Nir and D. Silverman, Nucl. Phys. B345 (1990) 301.}\
that if CKM unitarity were even moderately violated, then
New Physics contributions -- $t^\prime$-mediated box-diagrams
in models of a fourth quark generation and $Z$-mediated tree-diagrams
in models of non-sequential quarks --  would dominate
the mixing of neutral $B$-mesons. Consequently, either or
both of effects (b) and (c) are guaranteed to be much
more significant.

The upper bound on $|V_{td}|$ comes
from the assumption that the Standard Model contribution
saturates $x_d$. Therefore, its violation means that
effect (c) is important. We conclude that even if
$|V_{ts}/V_{td}|$ is outside of its Standard Model range,
it would not be the dominant source of violation for either
bound in \SMfinal.

2. In many extensions of the Standard Model, $R=R^{\rm SM}$
independently of whether there are significant new
contributions to neutral $B$ mixing. The most obvious
example is the Minimal Supersymmetric Standard Model (MSSM)
\ref\MSSM{S. Bertolini, F. Borzumati and A. Masiero, Phys. Lett. B194
(1987) 545; S. Bertolini, F. Borzumati, A. Masiero and G. Ridolfi,
Nucl. Phys. B353 (1991) 591; I.I. Bigi and F. Gabbiani, Nucl. Phys.
B352 (1991) 309.}:
$R^{\rm MSSM}=R^{\rm SM}$ is a result of the fact that the
mixing matrix for the gluino couplings to down quarks
and squarks is equal to the CKM matrix.  A second example
is multi-scalar doublet models with Natural Flavor Conservation
(NFC): $R^{\rm NFC}=R^{\rm SM}$ is a result of the fact that
the relevant charged scalar couplings are proportional, in most of
the parameter space, to $m_t V_{ij}$ where $V_{ij}$ is
the appropriate CKM element.

As all the considerations that lead to \SMRatio\ remain
valid in this class of models, the lower bound in \SMfinal\
remains valid, independent of whether the new contributions
to $x_s$ are significant.

On the other hand, the upper bound in \SMfinal\ does not
necessarily hold. If there are significant new contributions
(in this case to both $x_d$ and $x_s$), the upper bound is
relaxed to at least that of eq. \SMRatio.
Actually, with significant new contributions to $x_d$,
the lower bound on $|V_{td}|$ is relaxed to the CKM
unitarity bound: $|V_{ts}/V_{td}|\leq9$, leading to
$x_s\leq90$.

In some models, $R\geq R^{\rm SM}$. An example is a multi-scalar
doublet model with NFC where $|X|\geq{\cal O}(m_{H^\pm}/
\sqrt{m_b m_s})$
\ref\Grossman{Y. Grossman, WIS-94/3/Jan-PH (1994).}.
$X$ is a parameter
that arises from mixing of charged scalars and determines the size
of the lightest charged scalar Yukawa couplings that are
proportional to down-type masses. ($|X|$ can be
large enough only in models with more than two scalar doublets.)
In such models, again, the lower bound in \SMfinal\ holds,
but the upper bound could be significantly violated \Grossman.

In various other models, $R\approx R^{\rm SM}$ is a good
order of magnitude estimate. For example, in
multi-scalar models with no NFC but with horizontal symmetries
\ref\Hor{L.J. Hall and S. Weinberg, Phys. Rev. D48 (1993)
 R979; T.P. Cheng and M. Sher, Phys. Rev. D35 (1987) 3484;
A. Antaramian, L.J. Hall and A. Rasin, Phys. Rev. Lett.
 69 (1992) 1871; M. Leurer, Y. Nir and N. Seiberg, Nucl. Phys. B319
 (1993) 342.} one typically estimates
$R^{\rm Hor}\sim{m_s\over m_d}$ which is well within the range
of $R^{\rm SM}$. Another example is that of Extended
Technicolor (ETC) interactions that generate the top quark mass
\ref\ETC{L. Randall, Phys. Lett. B297 (1992) 309.}.
In these models we do not expect a strong violation of the
lower bound in \SMfinal, though it is not rigorously excluded.

Finally, there are models where the New Physics contribution
is much smaller than the Standard Model one. For example,
in Left-Right Symmetric models, $W_R$-mediated box-diagrams
are constrained to contribute less than about 15\% of the
Standard Model diagrams. Moreover,
the new contribution obeys $R^{\rm LRS}=R^{\rm SM}$:
this is a result of the fact that the mixing matrix
for $W_R$ couplings is similar to the CKM matrix.
(The situation could be different in models of $SU(2)_L\times
SU(2)_R\times U(1)_{B-L}$ gauge symmetry without the discrete
LRS and with fine-tuned mixing angles
\ref\LaSa{P. Langacker and S.U. Sankar, Phys. Rev. D40 (1989) 1569;
 D. London and D. Wyler, Phys. Lett. B297 (1992) 503.}.)
In supersymmetric models with quark--squark alignment (QSA)
\ref\QSA{Y. Nir and N. Seiberg, Phys. Lett. B309 (1993) 337;
M. Leurer, Y. Nir and N. Seiberg, WIS-93/93/Oct-PH (1993).}
the Supersymmetric diagrams contribute negligibly to $x_s$
and modify $x_d$ by no more than 15\%. In this type of models,
the Standard Model constraints on $x_s$ \SMfinal\ remain
essentially unchanged.

3. A third observation is that if $x_d$ is dominated
by the Standard Model contribution, then a violation of the
lower bound in \SMfinal\ is unlikely. The reason for that
is simple: if $x_d$ is accounted for by the $t$-mediated
box diagrams, then \SMRatio\ gives the {\it correct}
bounds on the Standard Model contribution to $x_s$.
 Therefore, in order
that the lower bound in \SMfinal\ is violated, the New Physics
has to interfere destructively with the Standard Model.
This requires that the two contributions are of the same
order of magnitude and of opposite signs. In the large
parameter space of New Physics models, such a possibility
usually requires fine-tuning.

4. The most interesting models, as far as near-future
measurements of $x_s$ are concerned, are those where large
contributions from New Physics to $x_d$ are possible and
where $R\neq R^{\rm SM}$. Is this case, the scaling from
the experimental value $x_d^{\rm exp}$ is misleading: the Standard
Model contributes $x_s^{\rm SM}=R^{\rm SM}x_d^{\rm SM}$
which could be smaller than the lower bound in \SMRatio\
(though not significantly smaller than the lower bound
in \SMdirect). This makes the search for $x_s$ in the range
$2\leq x_s\leq7$ very interesting: if $x_s$ is found
to lie in this range it will most likely imply that there
are significant new contributions to $x_d$! We next
describe two examples of such models.

\newsec{Models that Allow Small $x_s$}
1. Our first example is a model with extra
mirror down quarks, $D(3,1)_{-1/3}$ and $\bar D(\bar 3,1)_{+1/3}$.
Such particles are predicted by $E_6$ GUTs and in ``string
inspired'' frameworks. If the masses of these vector
quarks are not much larger than the electroweak breaking
scale, the $Z$-boson is likely to have non-negligible
flavor changing couplings to quarks $U_{ij}$. $Z$-mediated
tree diagrams will contribute to $x_s$:
\eqn\xsZ{
x_s^Z={\sqrt2 G_F\over6}\eta_{QCD}
(\tau_{B_s}m_{B_s})(B_{B_s}f_{B_s}^2)|U_{sb}|^2.}
The ratio between the new contributions to $x_s$ and to $x_d$,
\eqn\RZ{
R^{\rm Z}=\left({m_{B_s}\over m_{B_d}}{\tau_{B_s}\over\tau_{B_d}}
\right)\left({B_{B_s}f_{B_s}^2\over B_{B_d}f_{B_d}^2}\right)
\left|{U_{sb}\over U_{db}}\right|^2,}
could be very different from $R^{\rm SM}$. Moreover,
for $0.01\leq|U_{db}/V_{cb}|\leq0.04$, the $Z$-contribution
to $x_d$ is significant
\ref\Znisi{Y. Nir and D. Silverman, Phys. Rev. D42 (1990) 1477.}.
On the other hand, the experimental bound
on $BR(B\rightarrow X_s\mu^+\mu^-)$ gives $|U_{sb}/V_{cb}|\leq0.04$,
implying that the $Z$ contribution to $x_s$ is, at most,
25\% of the Standard Model contribution
\ref\Zsi{D. Silverman, Phys. Rev. D45 (1992) 1800;
Y. Nir, Lectures given in SSI-20, SLAC-PUB-5874 (1992);
G.C. Branco {\it et al.}, Phys. Rev. D48 (1993) 1167.}.
We conclude that in models with $Z$-mediated FCNCs,
$x_s$ is dominated by the Standard Model contribution
and the constraints replacing \SMfinal\ are
\eqn\Zfinal{
2\leq x_s\leq 50}
(where we have taken into account a possible 25\% effect
due to the $Z$ contribution).

2. The second example is a model with a fourth quark generation.
(Of course, four quark generation models are {\it not} a very
likely possibility in view of the LEP and SLC bounds on the number
of light left-handed neutrinos.)
Diagrams with one or two $t^\prime$ propagators replacing the
the Standard Model $t$ propagators contribute
\eqn\xsfour{\eqalign{
x_s^{\rm 4G}={G_F^2 m_W^2\over6\pi^2}&\eta_{QCD}
(\tau_{B_s}m_{B_s})(B_{B_s}f_{B_s}^2)\cr \times&
\left|2y_ty_{t^\prime}g_3(y_t,y_{t^\prime})
(V_{ts}^*V_{tb}V_{t^\prime s}^*V_{t^\prime b})+
y_{t^\prime}f_2(y_{t^\prime})(V_{t^\prime s}^*V_{t^\prime b})^2
\right|,\cr}}
where
\eqn\gthree{
g_3(y_i,y_j)=\left[{1\over4}-{3\over2(y_j-1)}-{3\over4(y_j-1)^2}
\right]{\ln y_j\over y_j-y_i}+(y_i\leftrightarrow y_j)-
{3\over4(y_i-1)(y_j-1)}.}
The bounds from $BR(B\rightarrow X_s\mu^+\mu^-)$
(see \ref\UAone{C. Albajar {\it et al.}, UA1 collaboration,
 Phys. Lett. B262 (1991) 163.}\
for the experimental bound and
\ref\HWS{W.-S. Hou, R.S. Willey and A. Soni, Phys. Rev. Lett. 58
(1987) 1608.}\ for the theoretical expression)
and from $BR(B\rightarrow X_s\gamma)$
(see \ref\CLEObsg{R. Ammar {\it et al.}, CLEO collaboration,
 Phys. Rev. Lett. 71 (1993) 674.}\
for the experimental bound and
\ref\HSS{W.-S. Hou, A. Soni and H. Steger, Phys. Lett. B192
(1987) 441.}\ for the theoretical expression)
are rather mild and allow
the new contributions to dominate $x_s$. As $R^{4G}\neq R^{\rm SM}$
and as $x_d$ may be dominated by $t^\prime$ contributions,
the lower bound on $x_s$ is relaxed.

To be more precise, we distinguish three cases:

(a) $\left|{V_{t^\prime s}^*V_{t^\prime b}\over V_{ts}^*V_{tb}}
\right|^2\ll{y_t\over y_{t^\prime}}$: the top contribution
dominates and the fourth generation induces small corrections
only. We have $3\leq x_s\leq 40$.

(b) $\left|{V_{t^\prime s}^*V_{t^\prime b}\over V_{ts}^*V_{tb}}
\right|^2\gg{y_t\over y_{t^\prime}}$: the $t^\prime$ contribution
dominates and $x_s$ could be significantly enhanced over its
Standard Model value. We have $3\leq x_s$ while the upper bound
could be ${\cal O}(10)$ weaker than in the Standard Model.

(c) $\left|{V_{t^\prime s}^*V_{t^\prime b}\over V_{ts}^*V_{tb}}
\right|^2\sim{y_t\over y_{t^\prime}}$: the $t$ and $t^\prime$
contributions are of the same order of magnitude. If the
relative phase between the two CKM combinations is real,
$x_s$ is enhanced. Only if  $\arg\left({V_{t^\prime s}^*
V_{t^\prime b}\over V_{ts}^*V_{tb}}\right)\sim\pi/2$
a significant destructive interference becomes possible.
Thus, to suppress $x_s$ below, say, 1 would require
fine-tuning of both the magnitude and the phase of
the mixing matrix. This confirms the results of ref.
\ref\London{D. London, Phys. Lett. B234 (1990) 354.}\
that finds that a small $x_s$ arises in only a tiny
region of the four generation model parameter space.
(For previous studies of $x_s$ in four generation models, see
\ref\HoSoxs{W.-S. Hou and A. Soni, Phys. Lett. B196 (1987) 9;
 J.L. Hewett and T.G. Rizzo, Mod. Phys. Lett. A3 (1988) 975.}.)

We conclude that in four generation models, if no fine-tuned
cancellations take place,
\eqn\FourFinal{
2\leq x_s^{4G}}
(where we allowed a reasonable destructive interference)
while the upper bound is high above the Standard Model bound.

\newsec{Conclusions}
Values of $x_s\leq7$ do not require fine-tuned cancellations
between Standard Model and New Physics contributions.
Instead, $2\leq x_s\leq7$ is possible under two conditions:
(a) There are significant new contributions to $x_d$; and
(b) The ratio of these contributions to $x_s$ and to $x_d$
is not proportional to $(V_{ts}/V_{td})^2$.
The types of New Physics most likely to fulfill these
conditions are extensions of the quark sector by
either sequential or non-sequential quarks.

The ``window'' that we find for naturally small $x_s$ depends on the
lower bounds on $m_t$ and $f_{B_s}$. For example, if experiments
find $m_t\geq160\ GeV$, the lower
bound on $x_s$ in eq. \SMdirect\ will change from 3 to 4; if lattice
calculations imply $f_{B_s}\geq0.19\ GeV$, the bound
will change to 5. (The window will be closed if $f_{B_s}\geq0.22\ GeV$
is established.)

We present the $x_s$ bounds
in various extensions of the Standard Model in Table 1.
The numbers presented in this Table are often a result
of a more detailed calculation than presented above.
For example, in the MSSM, we take into account that
supersymmetric diagrams may enhance the Standard Model
result by about 20\%, while in quark--squark alignment
models \QSA\ supersymmetric diagrams may modify $x_d$
in either direction by about 15\% and do not affect $x_s$.

\vskip 0.5cm
\centerline{Table 1}
\centerline{Bounds on $x_s$}
\vskip 0.5cm
\begintable
Model|\ SM\ |MSSM |QSA\QSA| NFC | Hor\Hor| ETC\ETC|\ LRS\ |$Z$-FCNC| 4 Gen \cr
$x_s\geq$|7|  7  |  6    | 7   |$\sim7$|$\sim7$| 7   | 2   | 2 \cr
$x_s\leq$|40| 50 | 40  |Large|$\sim90$|$\sim90$| 45  | 50  |Large
\endtable

 If experiments find $x_s<7$, it would have interesting implications
for CP asymmetries in neutral $B$ decays. As the likely explanation
of small $x_s$ is a large New Physics component in $x_d$, then
CP asymmetries in $B_d$ decays may differ significantly from the
Standard Model predictions. The combination of $x_s$ and CP asymmetry
measurements would be useful in closing in
on the source of deviations from the Standard Model.

\bigbreak
\centerline{{\bf Acknowledgements}}

I thank Lance Dixon, Ehud Duchovni,
Yuval Grossman, Miriam Leurer, Zoltan Ligeti and Jon Rosner
for useful discussions.  I am grateful to the Rutgers
Theory Group for their hospitality. YN is an incumbent of
the Ruth E.~Recu Career Development chair, and is supported in part
by the Israel Commission for Basic Research, by the United
States--Israel Binational Science Foundation (BSF), and by the
Minerva Foundation.

\listrefs
\end